# Privacy-preserving and Efficient Aggregation based on Blockchain for Power Grid Communications in Smart Communities

Zhitao Guan, Guanlin Si, Xiaosong Zhang*, Longfei Wu, Nadra Guizani, Xiaojiang Du, Yinglong Ma

**ABSTRACT**

Intelligence is one of the most important aspects in the development of our future communities. Ranging from smart home, smart building, to smart city, all these smart infrastructures must be supported by intelligent power supply. Smart grid is proposed to solve all challenges of future electricity supply. In smart grid, in order to realize optimal scheduling, a Smart Meter (SM) is installed at each home to collect the near real-time electricity consumption data, which can be used by the utilities to offer better smart home services. However, the near real-time data may disclose user's privacy. An adversary may track the application usage patterns by analyzing the user's electricity consumption profile. In this paper, we propose a privacy-preserving and efficient data aggregation scheme. We divide users into different groups and each group has a private blockchain to record its members' data. To preserve the inner privacy within a group, we use pseudonym to hide user's identity, and each user may create multiple pseudonyms and associate his/her data with different pseudonyms. In addition, the bloom filter is adopted for fast authentication. The analysis shows that the proposed scheme can meet the security requirements, and achieve a better performance than other popular methods.

## INTRODUCTION

Smart communities have many typical examples, such as smart home, smart building, and smart campus. Smart communities are built on top of smart power supply with smart grid. For instance, the implementation of smart home is very dependent on the Advanced Metering Infrastructure (AMI), which is one of the key technologies of smart grid. In AMI, two-way communication between the utility and the customers is achieved with the support of information and communication technologies. Smart meters (SM) are widely deployed to provide electric utilities with reliable power services. In order to realize optimal scheduling, large quantities of SMs have been installed in users' homes to collect near real-time electricity consumption data on a requested or scheduled basis [1]. Based on the data collected, the control center can draw the electricity consumption profile and then offer the dynamic pricing which allows users to benefit from by altering their electricity consumption behaviors. For example, some users will adjust their electricity consumption behaviors if they are aware of the high cost of electricity at peak demand times. Besides, a power plan can be developed in advance according to the total electricity consumption requests.

However, the near real-time electricity consumption data and the power consumption requests collected by SM may disclose user's privacy. With non-intrusive appliance load monitors (NALM), the adversary could track the application consumption patterns to infer user's behaviors, e.g., by analyzing the user's electricity consumption profile, which is a non-negligible threat to the user privacy. Besides, the power consumption requests used for developing the power plan may disclose user's activity plan in future. For example, a thief may find out when the user will not be at home from the power request, and commit theft by that time.

Although there have been many works contributing to the privacy preservation in smart grid, several challenges remain to be addressed. The first challenge is the availability of a trust party for data aggregation. Privacy is the most important concern when users submitting their sensitive data to a third party. In practice, it could be difficult to find an aggregator that is trusted by all users. The second challenge is to hide the connection between a user's real identity and the pseudonym. Pseudonym is a common solution to disguise user's identity. However, the connection between user's real identity and the pseudonym may be disclosed by matching the electricity consumption profile with user behaviors in a particular period of time. The third challenge is the authentication speed. As smart grid is a huge and complex system, conventional authentication mechanisms may not fit for smart grid considering the huge amount of computational overheads and time delays.

To tackle the above-mentioned challenges, we propose the following solutions. In lack of a trusted third party for data

*Xiaosong Zhang is the corresponding author.

Zhitao Guan, Guanlin Si, and Yinglong Ma are with School of Control and Computer Engineering, North China Electric Power University, Beijing, 102206, China (E-mail: guan@ncepu.edu.cn).

Xiaosong Zhang is with Center for Cyber Security, University of Electronic Science and Technology of China, Chengdu, China (E-mail: johnsonzxs@uestc.edu.cn).

Longfei Wu is with the Department of mathematics and Computer Science, Fayetteville State University, Fayetteville, NC, USA (E-mail: lwu@uncfsu.edu)

Nadra Guizani is with School of Electrical and Computer Engineering, Purdue University, West Lafayette, IN, USA (E-mail: nguizani@purdue.edu).

Xiaojiang Du is with the Department of Computer and Information Sciences, Temple University, Philadelphia, PA, USA (E-mail: dxj@ieee.org).

aggregation, one user is chosen randomly in each time slot to aggregate all users' data and record these data into the blockchain for message integrity. Since the blockchain is shared by all users in the same group, even if a malicious user is selected as the aggregator who tampers the record, the tampered data will be found by the others. To hide the connection between the user identity and pseudonym, we allow each user to create multiple pseudonyms and submit their electricity consumption data under different pseudonyms. To realize fast authentication while protecting user's privacy, we use bloom filter to judge the validity of pseudonyms and check the existence of fake pseudonym base on zero knowledge proof.

In this paper, we first introduce the related privacy-preserving techniques. Then, we give the preliminaries of our scheme design. Next, the proposed scheme is described in detail. After that, we conduct the security analysis and performance evaluation. At last, we conclude our paper.

## EXISTING PRIVACY-PRESERVING TECHNIQUES

Recently, various techniques have been proposed to address the problem of privacy preserving in smart grid. Reading through the related work, we summarize the related techniques from two aspects. One aspect focuses on protecting user's identity and the other aspect places emphasis on protecting user's data.

### A. Protecting user's identity

- **Privacy-preserving based on virtual ring**

A scheme based on virtual ring to protect user's identity is proposed in [2]. Users send their messages to the control center through the virtual ring, and the control center validates user's identity through ring signature without knowing the user's identity. The disadvantage of this kind of schemes is that if there is a malicious user sending falsified messages through the virtual ring, it is difficult to find out who the malicious user is because of the property of virtual ring.

- **Privacy-preserving techniques based on anonymity**

Anonymization [3] [4] is a common technique to protect user's identity. User's attributes can be classified into identity information, quasi-identifier and sensitive information. For an anonymity table, if the attributes have not been anonymized, the adversary may infer the relationship between the user's identity and the sensitive information by observing the quasi-identifiers such as age and gender. The limitation is that there must exist a reliable third party dedicated for processing the user's attributes. In our scheme, blockchain is adopted to achieve decentralized data processing.

- **Privacy-preserving based on pseudonym**

Pseudonym [5] is also a popular technique to protect user's identity. The registration process of pseudonym often involves ring signature, zero knowledge proof and so on. Gong *et al.* [6] propose a privacy-preserving scheme for incentive-based demand response in the smart grid, which adopts the discrete logarithm to create the pseudonym and hides the user identity through ring signature during the registration process of pseudonym. To hide this relationship, in our scheme, each user is allowed to adopt multiple pseudonyms during data collection.

### B. Protecting user's data

- **Privacy-preserving based on household battery**

A scheme using the battery to hide the near real-time data is proposed in [7] [8]. When the household consumption profile goes high, the battery discharges. Otherwise, it charges. Therefore, user's electricity consumption profile can be balanced by the house battery and the user's privacy can be protected very well. However, the disadvantage of this kind of scheme is that using battery to hide user's electricity consumption profile may conflict with user's economic interest, especially battery charging at the peak times. Besides, the effect of privacy preserving depends on the capacity of battery. The tradeoff between privacy and economic cost is analyzed in [9].

- **Privacy-preserving based on data aggregation**

Data aggregation is a common scheme to protect user's data and it contains homomorphic encryption and data obfuscation. The homomorphic encryption allows the intermediary agent to operate on the encrypted data with no information about the plaintext. Paillier encryption and BGN encryption are two typical homomorphic encryption methods, and are often used to calculate the sum of electricity consumption data based on the property of additive homomorphism. The main idea of data obfuscation is to add noise into the original data to obfuscate user's electricity consumption data. Based on the noise being added, it is divided into two categories: random data obfuscation [10] and non-random data obfuscation [11]. In the former category, the noise is completely random. In the latter category, each user's noise is usually assigned by the key management center to ensure that the summation of noise is equal to zero.

- **Privacy-preserving authentication techniques**

Privacy preserving and authentication are two related topics [12] [13]. Cheung *et al.* [14] propose a credential-based privacy-preserving scheme based on the blind signature. A consumer generates lots of credentials and asks the control center to sign these credentials blindly. The credentials will be sent to the control center for proof of user's identity when the user asks for more power. However, the computational overhead of this scheme is heavy since the control center need to sign numerous credentials each month.

## PRELIMINARIES

### A. Blockchain

Blockchain is a chronologically ordered chain of blocks which can be considered as a replicated state machine. For each block, the transactions are hashed in Merkle Tree. The root hash of Merkle Tree and the hash of previous block are recorded in the block header. As the blocks are connected by the hash of the previous block, anyone who wants to change the backdated transactions has to modify that block and all the following blocks, which is considered difficult.

To generate a valid block, Proof-of-Work, a puzzle that is difficult to solve but easy to verify, is adopted to select a suitable user as the mining node. Then, the mining node broadcasts the completed block and the other nodes can check the validity of the new block based on the Proof-of-Work. However, Proof-of-Work usually needs to spend huge amount of computational cost. For instance, in Bitcoin, to generate a valid block, mining node must keep on trying different random number until the hash of the block conforms to the regulations.

As the utility terminal devices only support lightweight operations, we design a simple solution to randomly select the mining node. The user whose data is the closest to the average electricity consumption data in a time slot will be selected as the mining node. Because no one can confirm the value of average electricity consumption data before receiving all the users' data, the randomness and uniqueness of the selection of mining node can be guaranteed.

**B. Bloom Filter**

Bloom filter is a fast query algorithm that can be used for fast authentication. It can quickly confirm whether an ID is legal in the system. The authentication process can be divided into two stages.

- **Initialization of bloom filter**

For an array of $\theta$ elements, we initially set each element to be zero. Given an ID, we choose k hash functions to calculate k hash values of this ID as the index values and set the mapping values of the corresponding positions in the array to be one. The index value in the array is calculated as $h_i(ID) \mod \theta$. Note that the mapping value will not increase if multiple hash values refer to the same position.

- **Authentication by bloom filter**

To authenticate the validity of an ID, we can calculate $k$ hash values of the ID with the k hash functions to obtain the index values.

If the mapping values of the ID satisfy that each of the index value is equal to $h_i(ID) \mod \theta$ and contains zero in the bloom filter, this ID is illegal. As shown in Fig.1, $ID_1$ and $ID_2$ come from two valid users. $ID_3$ is fabricated by an adversary. By using the bloom filter, the illegal $ID_3$ can be detected easily based on the fact that the mapping values of $ID_3$ contains zero.

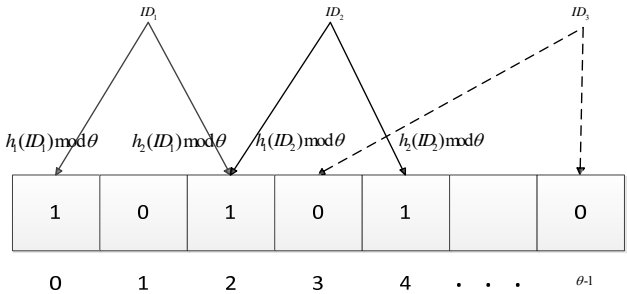

Figure.1 Bloom filter

**C. Zero knowledge proof**

Zero knowledge proof means that the verifier can prove the correctness of the input but has no information about the content of the input.

In this paper, we take user's pseudonym as our example. $pk$ and $sk$ represent a user's public and private key, respectively. We consider the public key $pk$ as a user's pseudonym and no one knows this user's real identity. Then, the user sends $EK_{sk}(m|pk)$ to the verifier. Here, $EK_{sk}(m|pk)$ represents the concatenation of message $m$ and pseudonym $pk$ encrypted by the private key $sk$. The verifier can confirm if the user's pseudonym is real by decrypting $EK_{sk}(m|pk)$ with the user's public key, but have no information about the user's real identity.

## PRIVACY-PRESERVING DATA AGGREGATION SCHEME

**A. System model**

As shown in Fig.2, the system model is a multi-tier smart grid communication network that consists of neighborhood area networks (NAN) and wide area networks (WAN). The NAN is formed by a large number of smart meters (SM) in the neighborhood. SMs send their meter readings to the mining node for data aggregation. The aggregated data of each group are sent to the center unit through the WAN.

We divide users into different groups according to their electricity consumption type. In each time slot, a user is chosen as the mining node according to the average electricity consumption data. The mining node is responsible for aggregating the data and recording these data into the private blockchain.

Key management center (KMC) is mainly responsible for initializing all of the keys for users. It generates multiple public and private keys for each user and takes the public key as the user's pseudonym. Then, it creates a bloom filter for each group by collecting the pseudonyms and sends the bloom filter to all users in the corresponding group.

The aggregated data of each group will be sent to the center unit through the wide area network. The control center can draw the electricity consumption profile based on the aggregated near real-time data for power planning and dynamic pricing.

The billing center is responsible for calculating each user's billing data based on the blockchain of each group when the billing date comes.

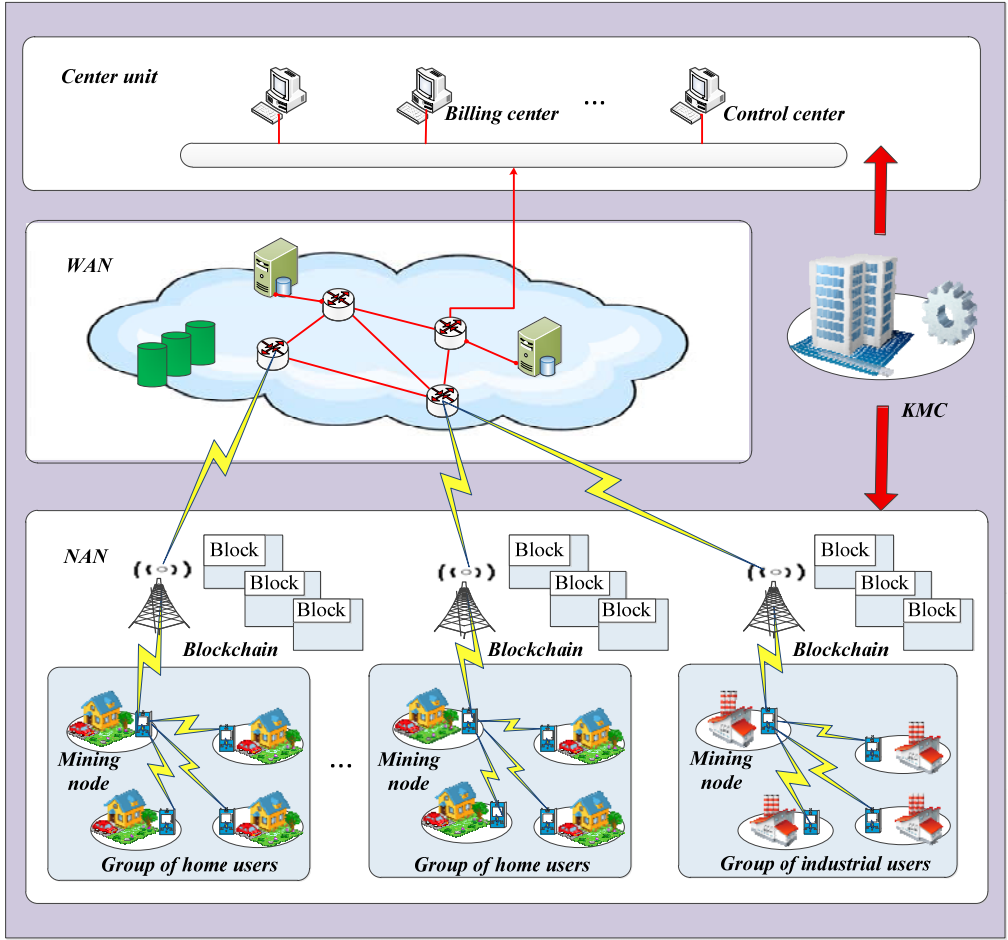

Figure.2 System model

**B. System initialization**

Fig.3 shows the architecture of our scheme in detail. KMC is responsible for bootstrapping the whole system in the beginning. Specifically, it executes the following procedures for system initialization

● **Key generation**

RSA algorithm: KMC chooses two large primes $p$, $q$, and calculates n = $pq$. Then, it chooses an integer $e$ which satisfies $1 < e < \varphi(n)$. Here, $\varphi(n)$ denotes the Euler function and the great common divisor of $e$ and $\varphi(n)$ is equal to one. Next, KMC calculates a parameter $d$ which satisfies $de = 1(\mod n)$. $e$ represents the private key and $d$ represents the public key.

Each user sends his ID to KMC for registration and acquires the public and private keys by running the RSA algorithm. Note that each user can acquire multiple pairs of public and private keys and takes the public keys as their pseudonyms.

● **Bloom filter**

Users are divided into different groups according to their electricity consumption types. Then, KMC creates the bloom filter for each group.

First, KMC sets an array which has $\theta$ bits. Then, it uses $k$ hash functions to calculate the hash values of all the pseudonyms in the same group. The mapping value whose index value is equal to $h_i(pk) \mod \theta$ is set to one. At last, KMC sends the bloom filter to all the users in the same group.

**C. Data collection**

To hide the identity from other users in the same group, we use pseudonyms to replace the user's true identity and each user can bind his electricity consumption data with multiple pseudonyms to achieve further obfuscation.

● **Data allocation**

In each time slot, such as 15min, each user randomly splits his/her electricity consumption data to be associated with different pseudonyms and publishes his electricity consumption data $m_i$, pseudonym $pk_i$ and $EK_{sk_i}(m_i | t | pk_i)$ in the group. Here, $EK_{sk_i}(m_i | t | pk_i)$ represents the concatenation of $m_i$, timestamp $t$ and pseudonym $pk_i$, encrypted by the sender's private key $sk_i$.

It's noteworthy that the allocated data can be negative. Therefore, user's privacy-sensitive data can be protected during the off-peak time, such as midnight. Besides, each user can assign his electricity consumption data into more or less pseudonyms as needed. At last, the user will publish $\{m_i, t, pk_i, EK_{sk_i}(m_i | t | pk_i)\}$ in the group.

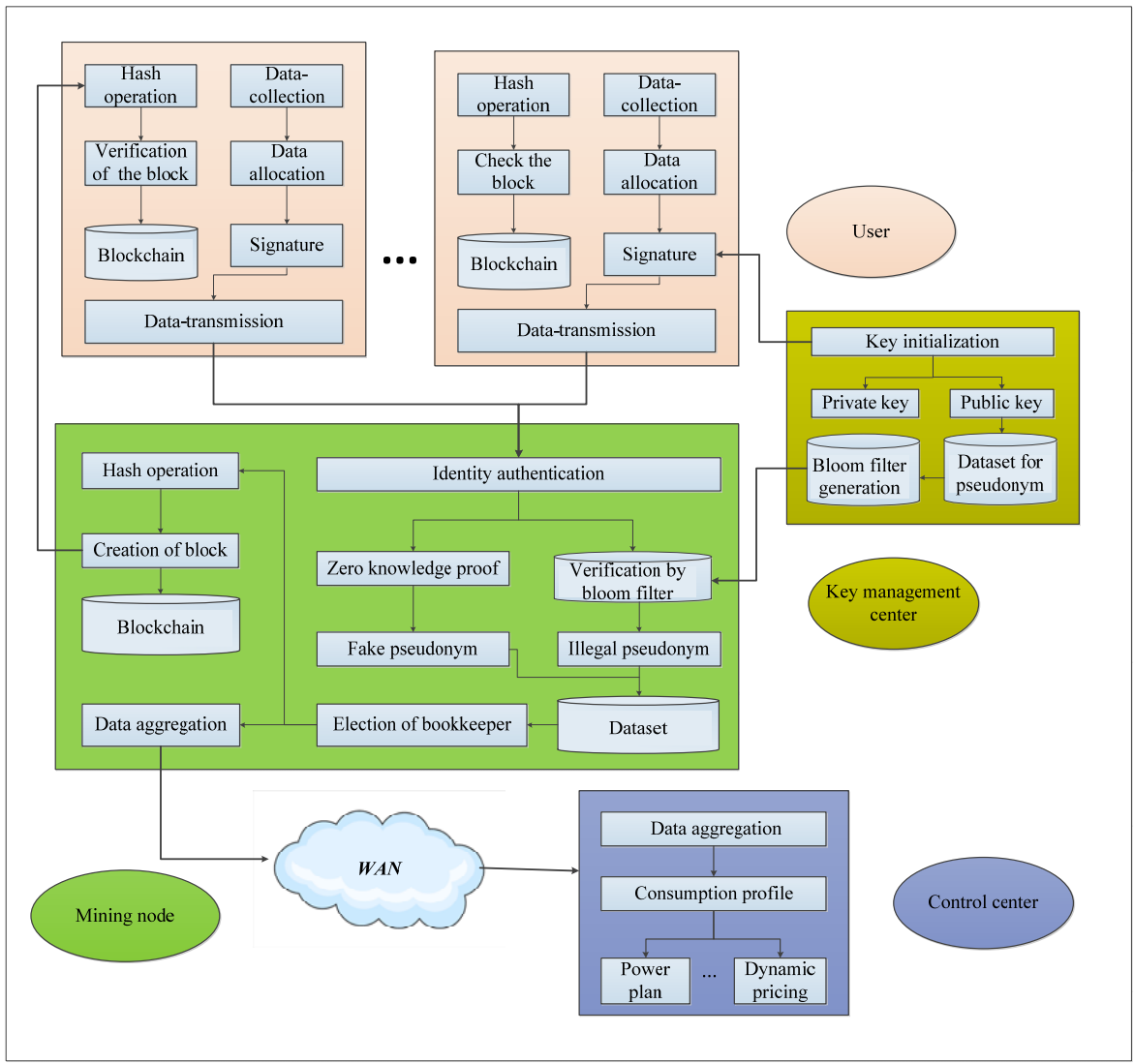

Figure.3 The architecture of our scheme

### D. Identity-authentication

- **Authenticity of user's pseudonym**

In case that an adversary forges legal a user's pseudonym, we verify the authenticity of user pseudonyms by zero knowledge proof.

When a user receives $\{m_i, t, pk_i, EK_{sk_i}(m_i | t | pk_i)\}$ sent by another user, he uses the sender's pseudonym $pk_i$ to check the signature $EK_{sk_i}(m_i | t | pk_i)$. If $EK_{sk_i}(m_i | t | pk_i)$ can be decrypted by $pk_i$ correctly, the pseudonym is proved to be authentic.

- **Validity of user's pseudonym**

In case that an unregistered adversary constructs a pseudonym and sends false data to the system, bloom filter is used to verify the validity of user's pseudonym.

For each received pseudonym, the receiver will use k hash functions to calculate the index values of this pseudonym as $h_i(ID) \bmod \theta$. If all the mapping values don't contain zero, the pseudonym is proved to be legal. Otherwise, the pseudonym will be considered as illegal and dropped by the receiver.

### E. Selection of mining node

After the identity authentication, all users will determine the mining node to aggregate their electricity consumption data and record these data into the blockchain.

Firstly, each user calculates the average electricity consumption data based on all the received data. Secondly, the one whose data is closest to the average is selected as the mining node.

There may be multiple pseudonyms whose average electricity consumption data have the same distance to the average, which means all of them will be the mining nodes in this time slot. However, such particular case doesn't affect the generation of new block, because the new block generated by either mining node is the same. No one knows the average before collecting data from all users in the group, therefore, it is difficult for a malicious user to increase the probability of being selected as a mining node.

### F. Creation of new block

The structure of a block is shown in Fig.4. After the mining node is selected, the electricity consumption data will be recorded into the blockchain and published among all users in this group for message authentication.

Firstly, the electricity consumption data are hashed by the mining node in the Merkle Tree. Secondly, the mining node records the root hash, timestamp, hash of the previous block, pseudonym and the average into the block header. At last, the new block will be published to the other users in the group for the message authentication.

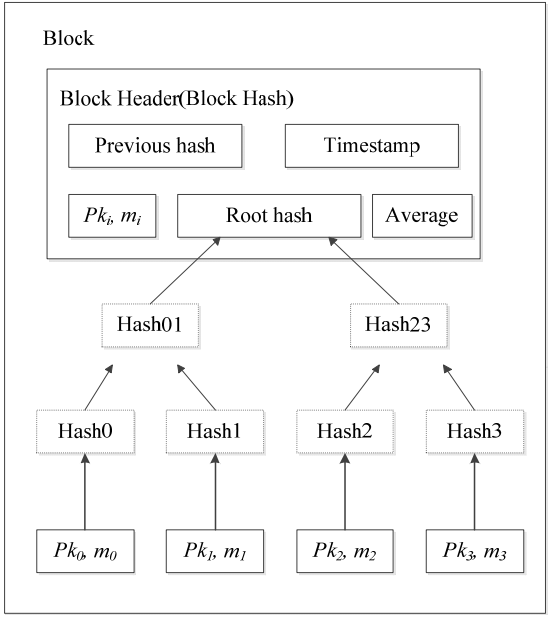

Figure.4 The structure of block in our scheme

### G. verification of the block

After receiving the new block, each user starts to verify the authenticity of the records. If the records in the new block are right, the user will attach this block to the blockchain which is saved in his dataset. If there is no one against the records in the new block, the mining node will send the sum of electricity consumption data to the control center.

To satisfy the real-time dispatch in the smart grid, each user only needs to verify the records related to his/her own data and the verification can be carried out offline. Therefore, we can aggregate the near real-time data with blockchain in a decentralized manner.

### H. Dispatching and billing

After receiving the sum of electricity consumption data from each group, the control center will draw the electricity consumption profile and offer the dynamic pricing to encourage users to adjust their electricity consumption behaviors.

In each billing cycle, such as a week or a month, the private blockchain of each group will be sent to the billing center. With the dynamic pricing in different times denoted by $p_0, p_1, ... p_f$, the billing center can calculate the electricity bill denoted by $P_{sum}$ according to the following formula: $P_{sum} = E_0 p_0 + E_1 p_1 + ... E_f p_f$, where $E_0, E_1, ... E_f$ denote the electricity consumption data in different time slots recorded in blockchain. As blockchain can guarantee the data integrity, we can realize the accurate billing easily.

## SECURITY ANALYSIS

### A. Data Privacy

For an adversary out of the group, he/she only knows the sum of electricity consumption data from the mining node. Therefore, the individual private data will not be disclosed. The mining node is chosen according to the average electricity consumption data in each time slot no one can guess the accurate average before the entire users' data in the group are collected. Therefore, even if an insider adversary exists, his/her chance of becoming the mining node is no larger than other users. Besides, we use blockchain to record user's data and the new block generated by the mining node will be verified by all the users in the group.

### B. Authentication

The authentication problem can be divided into two aspects in our scheme. One is to verify the authenticity of user's pseudonym in case that an adversary forges the legal user's pseudonym. The other one is to check the validity of user's pseudonym in case that an unregistered adversary sends false data to the control center.

For the authenticity of user's pseudonym, we take the public key as the pseudonym and use the asymmetric encryption to provide the zero knowledge proof. If an adversary tries to forge a legal user's pseudonym, he will fail due to the lack of the private key.

For the validity of user's pseudonym, we use bloom filter to quickly check whether the user's pseudonym has been registered. If an adversary tries to construct a pseudonym without registering at KMC, he will fail since the mapping values in bloom filter contain at least one zero.

In fact, bloom filter has error probability. As hash collision may happen during the hashing operation, an unregistered adversary may construct a pseudonym to pass through the bloom filter, which means all the mapping values in the bloom filter equal to one. However, the probability of hash collision is extremely low and the error probability is related to the amount of users, the number of hash functions and the length of the bloom filter. Therefore, if we set the array size sufficiently large, the probability of hash collision will decrease and the error probability can be ignored.

## PERFORMANCE EVALUATION

We evaluate our scheme in time complexity by comparing with the conventional authentication scheme that does not adopt the bloom filter. Here, we set the error probability of the bloom filter as 0.01, suppose that the amount of users in a group ranges from 0 to 200.

As shown in Fig.5, the time complexity of our scheme is much lower than the conventional scheme without bloom filter during the identity-authentication process.

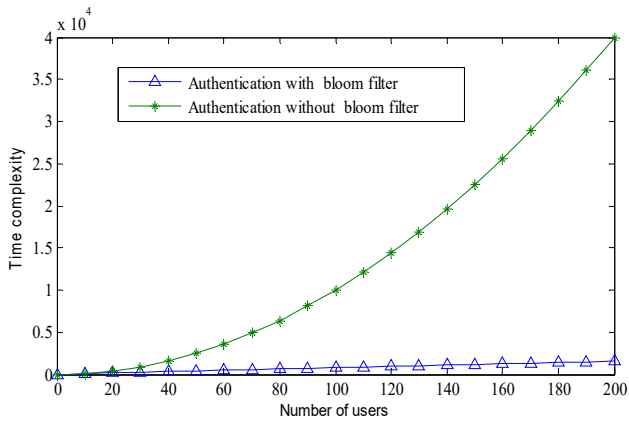

Figure.5.The time complexity about authentication

We compare our scheme with PPM-HAD [14] and DG-APED [15] in computational cost. As Fig.6 shows, our scheme has less computational cost than the other two schemes.

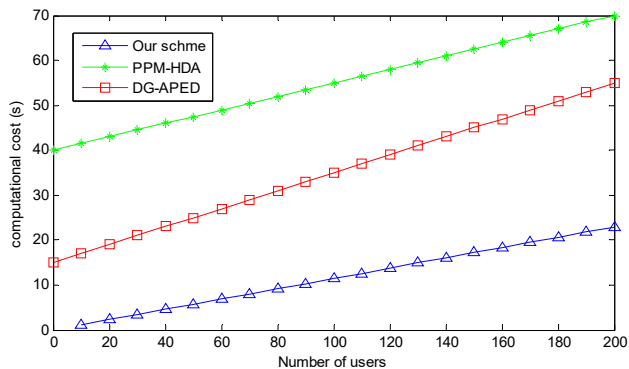

Figure.6.The computational cost

## CONCLUSION

We propose a privacy-preserving and efficient data aggregation scheme based on the blockchain to preserve user's privacy in smart grid. In each time slot, we select the mining node based on the difference between the user's electricity consumption data and the average value to guarantee the fairness of the selection. Then, the mining node records all users' data into the blockchain and publishes the blockchain within the group to ensure the message integrity. To defend against the snoop from other users in the same group, we adopt pseudonyms to protect user's identity and each user associate his data with multiple pseudonyms for further obfuscation. What's more, bloom filter is adopted to realize the fast authentication. The performance evaluation shows that our scheme has significant advantage compared with the conventional authentication scheme and other popular data aggregation schemes in computational efficiency. In the future work, we will continue improving our scheme to reduce the computational overhead caused by the authentication especially during the system initialization.

## ACKNOWLEDGMENT

This work is partially supported by Natural Science Foundation of China under grant 61402171, the Fundamental Research Funds for the Central Universities under grant 2015ZD12, 2016MS29.

## BIOGRAPHIES

**Zhitao Guan** is currently an Associate Professor at the School of Control and Computer Engineering, North China Electric Power University (NCEPU). He received his BEng degree and PhD in Computer Application from Beijing Institute of Technology (BIT), China, in 2002 and 2008, respectively. His current research focuses on smart grid security, wireless security and cloud security. Dr. Guan has authored over 20 peer-reviewed journal and conference papers in these areas.

**Guanlin Si** is currently an Master candidate at the School of Control and Computer Engineering, North China Electric Power University (NCEPU). He received his BEng degree from Hebei Agricultual University in 2015. His current research focuses on smart grid security.

**Xiaosong Zhang** is a pofessor of School of Computer Science and Engineering,University of Electronic Science and Technology of China(UESTC),ChengDu, China. His research interests are in network and information technology security and applicaiton,etc, his work has been published in a number of different journals. He is also the distinguished professor of Yangtze River and has lead a series of national science project in recent years.

**Longfei Wu** obtained his Ph.D. degree in computer and information sciences from Temple University in July 2017, advised by Dr. Xiaojiang Du. He received his B.E. degree in telecommunication engineering from Beijing University of Posts and Telecommunications in July 2012. He is currently an assistant professor in the Department of Mathematics and Computer Science at Fayetteville State University. His research interests include the security and privacy issues in mobile devices, implantable medical devices, and Internet-of-Things.

**Nadra Guizani** is a lecturer of Computer Science at Gonzaga University. She is currently a PhD student at Purdue University completing a thesis in Prediction and Access Control of Disease Spread Data on Dynamic Network Topologies. Research interests include machine learning, mobile networking, large data analysis and prediction techniques. Active member in Eta Kappa Nu Honors Society, Women in Engineering Program, and Computing Research Association for Women.

**Xiaojiang Du** is a professor in the Department of Computer and Information Sciences at Temple University, USA. Dr. Du received his B.S. Tsinghua University, Beijing, China in 1996, and his Ph.D. degree from the University of Maryland College Park in 2003. His research interests are wireless networks, security, and systems. He has authored over 230 journal and conference papers in these areas. Dr. Du has been awarded more than $5 million US dollars research grants.

**Yinglong Ma** is a professor with the School of Control and Computer Engineering, North China Electric Power University, Beijing, China. He received his Ph.D degree from the Institute of Software, Chinese Academy of Sciences, Peking, China, in 2006. His main research areas include Knowledge Engineering, Big Data Analysis and Information Forensics. He has published more than 60 scientific papers in some international/domestic journals and conferences in the last decades.

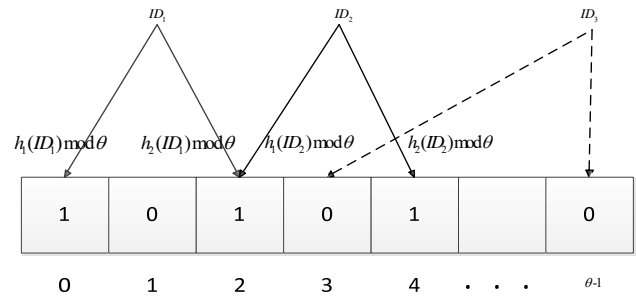

Figure.1. Bloom filter

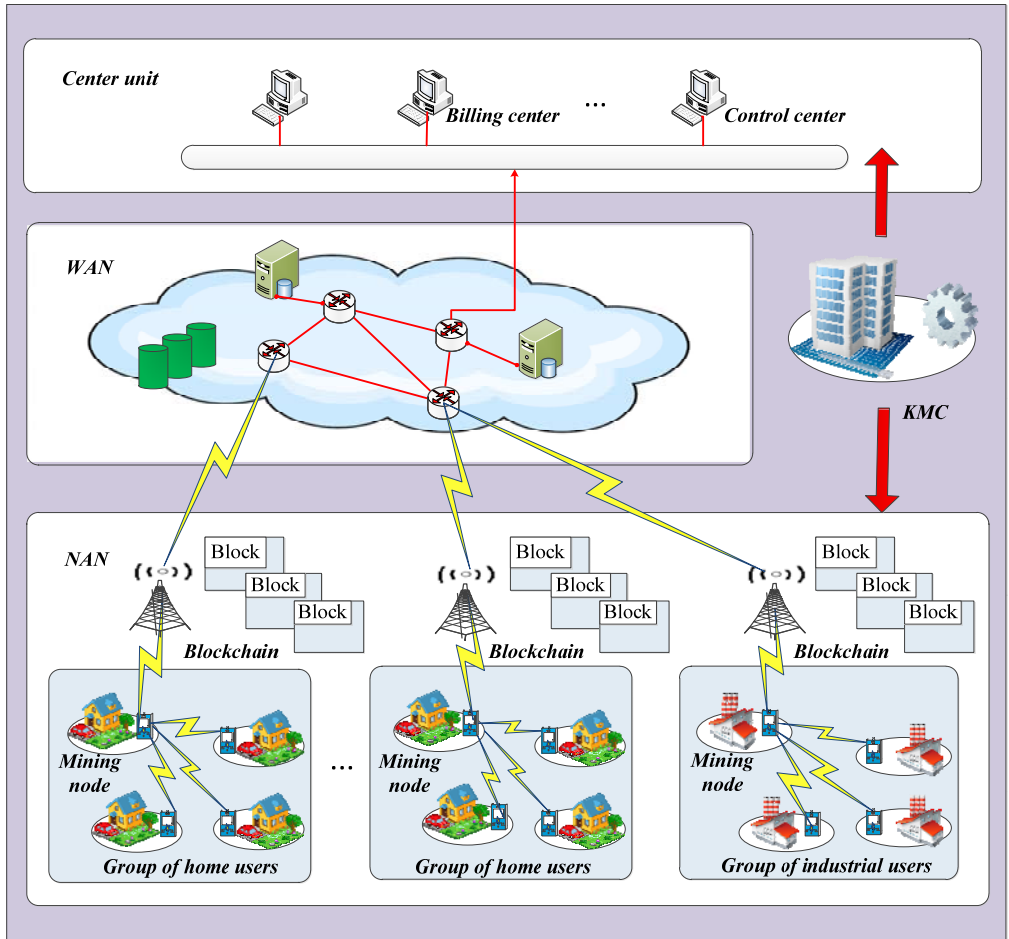

Figure.2. System model

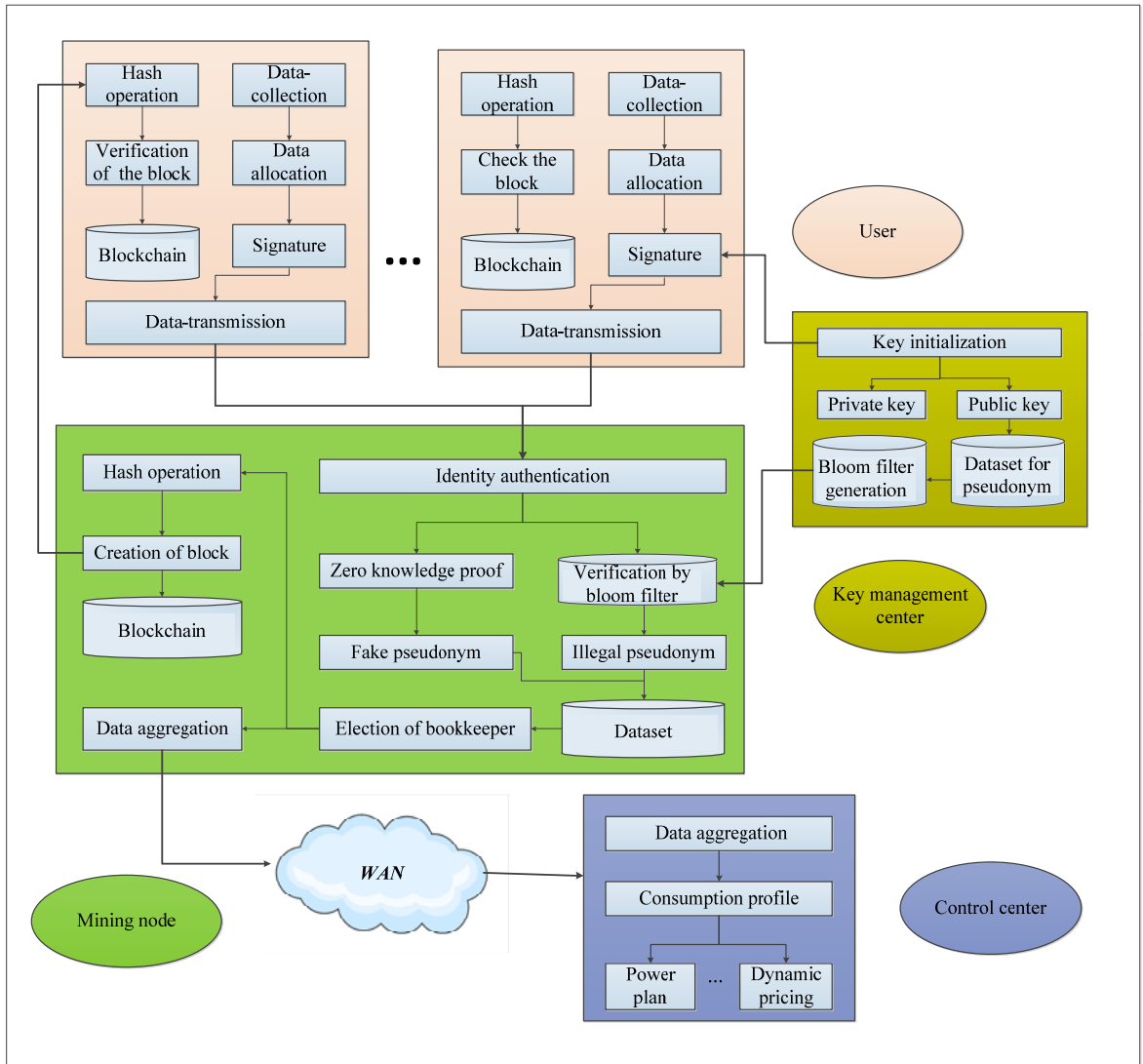

Figure.3. The architecture of our scheme

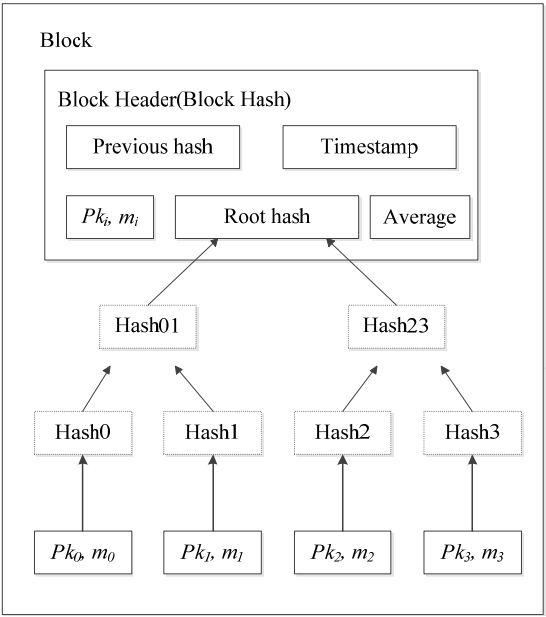

Figure.4. The structure of block in our scheme

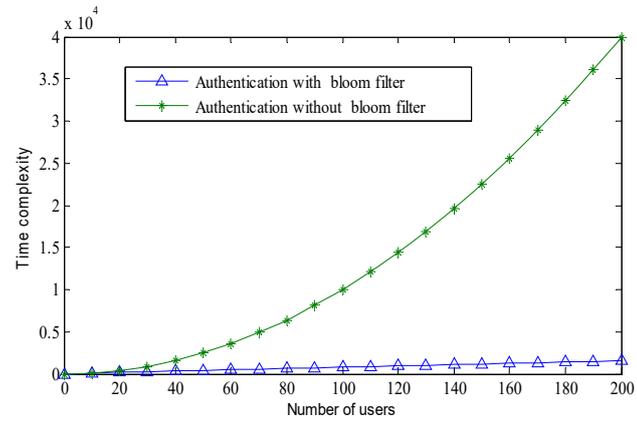

Figure.5. The time complexity about authentication

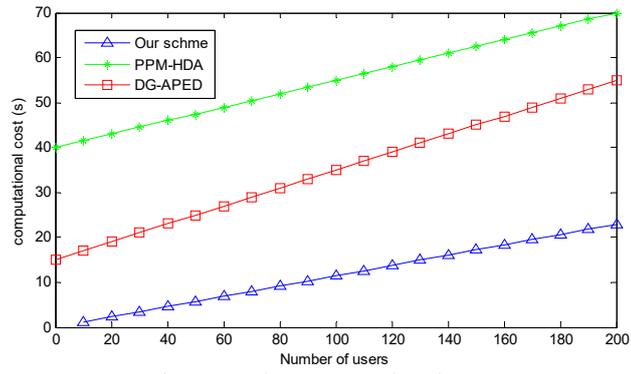

Figure.6. The computational cost